\documentclass[aps,preprint,prd,nofootinbib]{revtex4}
\usepackage{graphicx}
\usepackage{psfrag}

\usepackage{subfigure}
\usepackage{array}
\usepackage{graphicx}
\usepackage{amsmath}
\usepackage{amssymb}
\usepackage{mathrsfs}
\usepackage{bm}
\usepackage{slashed}
\usepackage{threeparttable}
\usepackage{multirow}
\usepackage[colorlinks, linkcolor=black, anchorcolor=black, citecolor=black]{hyperref}
\usepackage[amsmath,thmmarks,hyperref]{ntheorem}
\usepackage{soul}

\allowdisplaybreaks[4]

\begin{document}

\title{Doubly-charged scalar in four-body decays of neutral flavored mesons}
\author{Tianhong Wang\footnote{thwang@hit.edu.cn}, Geng Li\footnote{karlisle@hit.edu.cn}, Yue Jiang\footnote{jiangure@hit.edu.cn}~and~Guo-Li Wang\footnote{gl\_wang@hit.edu.cn}\\}
\address{Department of Physics, Harbin Institute of Technology, Harbin, 150001, China}

\baselineskip=20pt

\begin{abstract}

In this paper, we study the four-body decay processes of neutral flavored mesons, including $\bar K^0$, $D^0$, $\bar B^0$, and $\bar B_s^0$. These processes, which induced by a hypothetical doubly-charged scalar particle, violate the lepton number. The quantity $Br\times\left(\frac{s_\Delta h_{ij}}{m_\Delta^2}\right)^{-2}$ of different channels are calculated, where $s_\Delta$, $h_{ij}$, and $M_\Delta$ are parameters related to the doubly-charged scalar. For $\bar K^0\rightarrow h_1^+h_2^+l_1^-l_2^-$, $D^0\rightarrow h_1^-h_2^-l_1^+l_2^+$, and $\bar B_{d,s}^0\rightarrow h_1^+h_2^+l_1^-l_2^-$, it is of the order of $10^{-13}\sim 10^{-11}$ ${\rm GeV^4}$, $10^{-17}\sim 10^{-10}$ ${\rm GeV^4}$, and $10^{-17}\sim 10^{-10}$ ${\rm GeV^4}$, respectively. Based on the experimental results for the $D^0\rightarrow h_1^-h_2^-l_1^+l_2^+$ channels, we also set the upper limit for $\frac{s_\Delta h_{ij}}{M_\Delta^2}$.  
\end{abstract}

\maketitle

\section{Introduction}

In the previous work~\cite{wang18}, we have studied the lepton number violation decays of the $B_c^-$ meson induced by a doubly-charged Higgs boson. This particle generally appears in the left-right symmetric models~\cite{pati, Moha, Sen} and in the Type-II see-saw models~\cite{Magg, Laz, Moh81, Cheng} specifically. As it can decay into two leptons with the same charge, which indicates the lepton number violation, such processes of top quark, $\tau^-$~\cite{quin13}, and charged mesons, such as $K^-$, $D^-$, $D_s^-$, $B^-$~\cite{ma09, bam15, chak16, pic97} induced by this particle have been investigated extensively. As the lower bound of the mass of the doubly-doubly charged Higgs boson is around 800 GeV~\cite{ATLAS, CMS}, these low energy processes have extremely small branching ratios. Although it is not likely these channels can be detected recently, as experiments collect more and more data,  the upper limit of the branching ratios for such decay processes will be set more and more stringently (for $K$, it is of the order of $10^{-10}$ in PDG~\cite{pdg}). One can also use them to give some constraints on the effective short-range interactions~\cite{quin17}.

In Ref.~\cite{wang18} we considered both the three-body and four-body decay channels of $B_c^-$ meson. These channels violate the lepton number. In this paper, we investigate the doubly-charged Higgs boson induced lepton number violation processes of the neutral flavored mesons. Different from the charged meson case, where the annihilation-type diagram and the two $W$ meson emitting diagram contribute to the amplitude, for the neutral meson, the light antiquark just keeps as a spectator (see Fig.~1). Theoretically, this will make the calculation more simpler, as there is no complexity brought by the cascade decay. In the decay products, two leptons have the same charge, and so do two mesons. These decay modes have no background in the standard model, which makes them also interesting experimentally. 

\begin{figure}[ht]
\centering
\renewcommand{\thesubfigure}{(\Alph{subfigure})}
\subfigure[]{\includegraphics[scale=0.39]{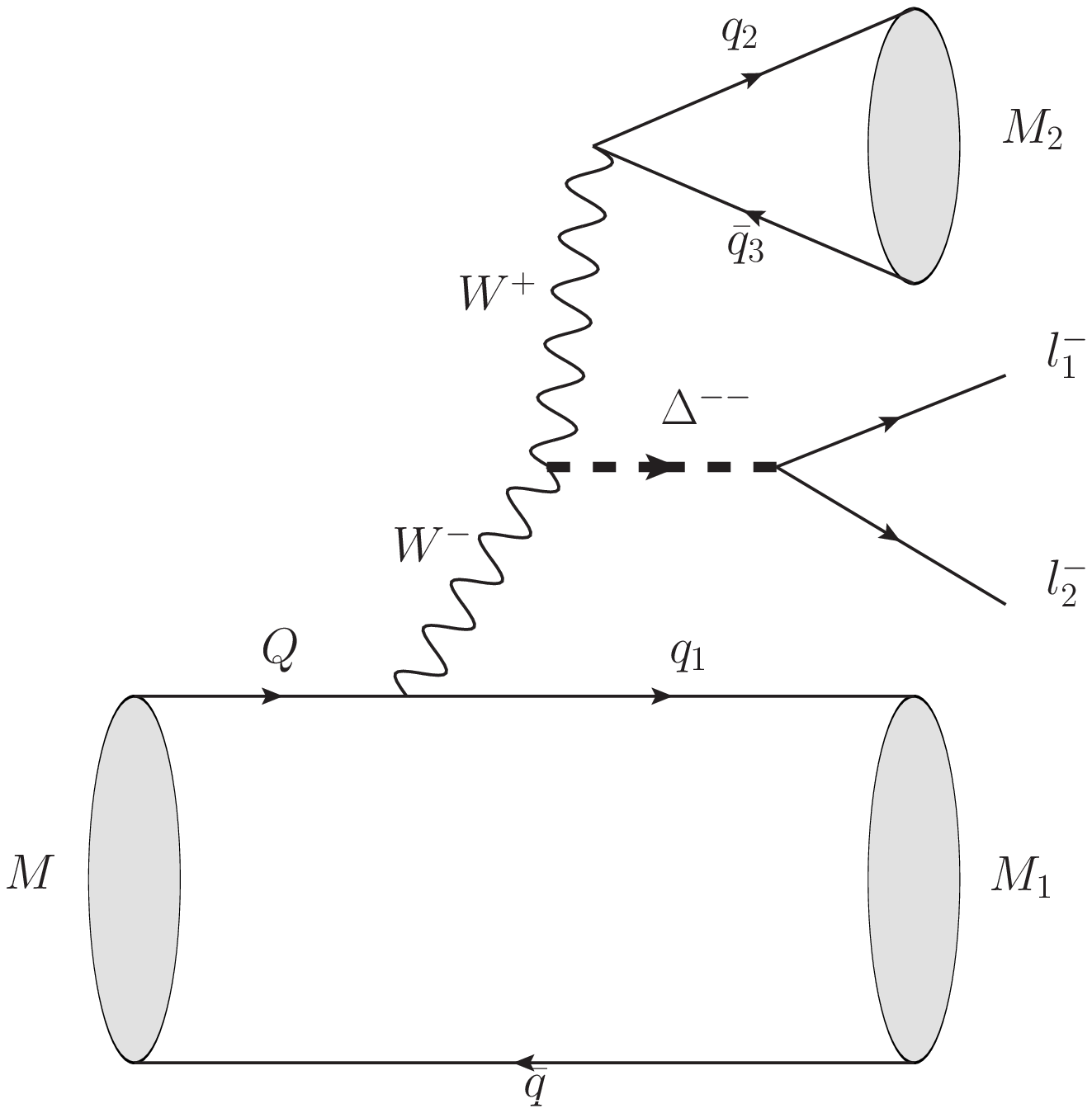}}
\hspace{2cm}
\subfigure[]{\includegraphics[scale=0.39]{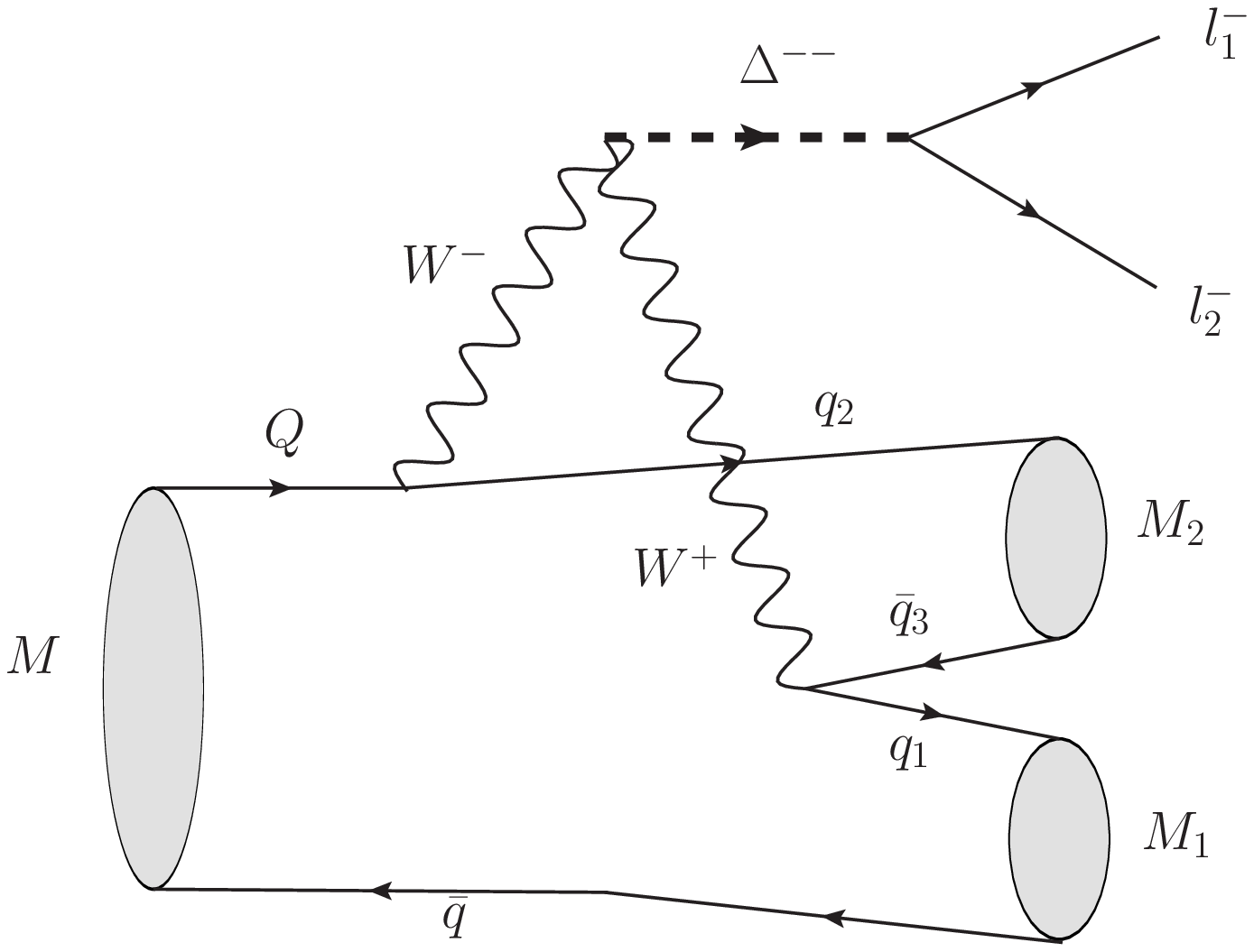}}
\hspace{2cm}
\subfigure[]{\includegraphics[scale=0.39]{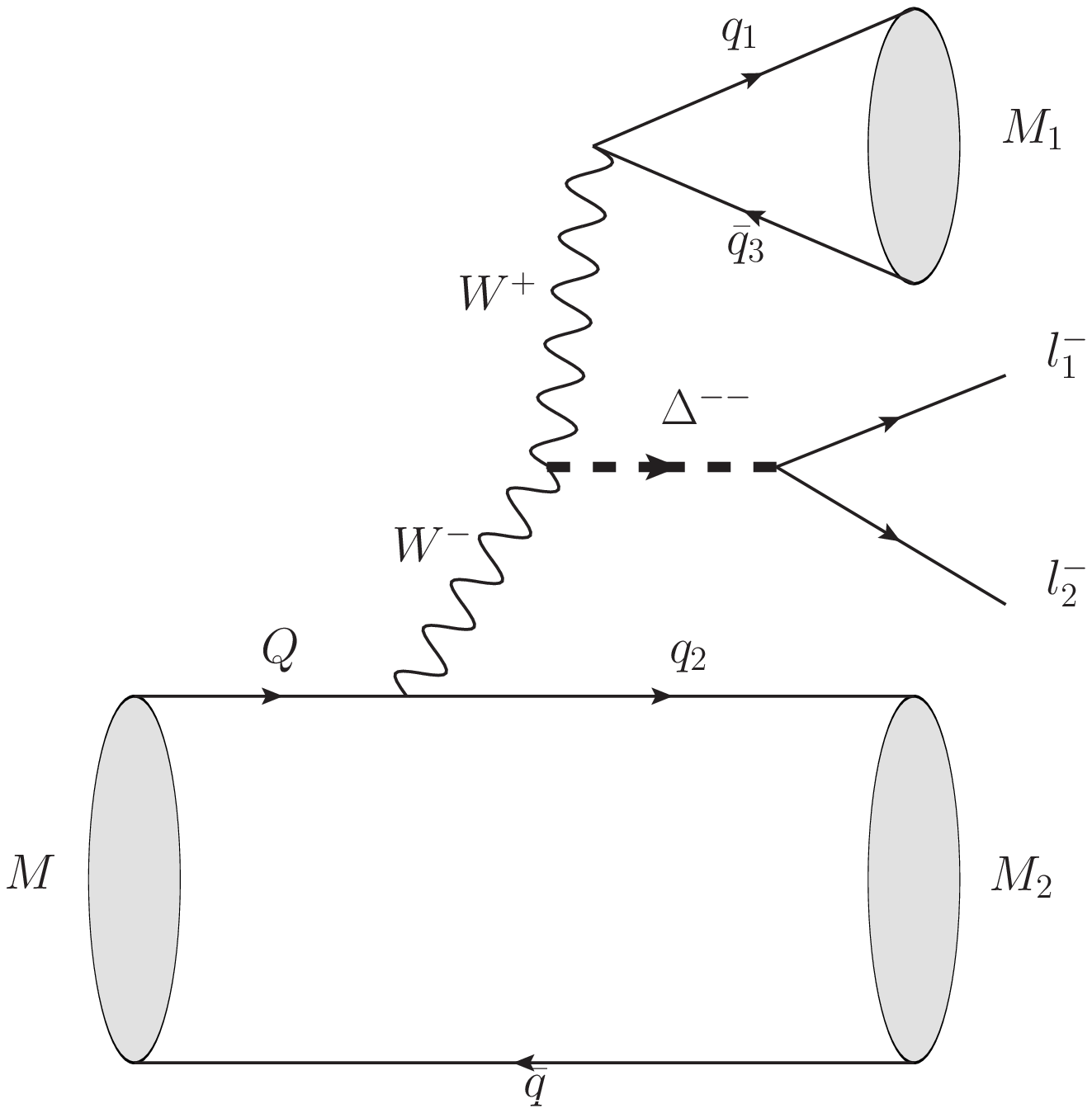}}
\hspace{2cm}
\subfigure[]{\includegraphics[scale=0.39]{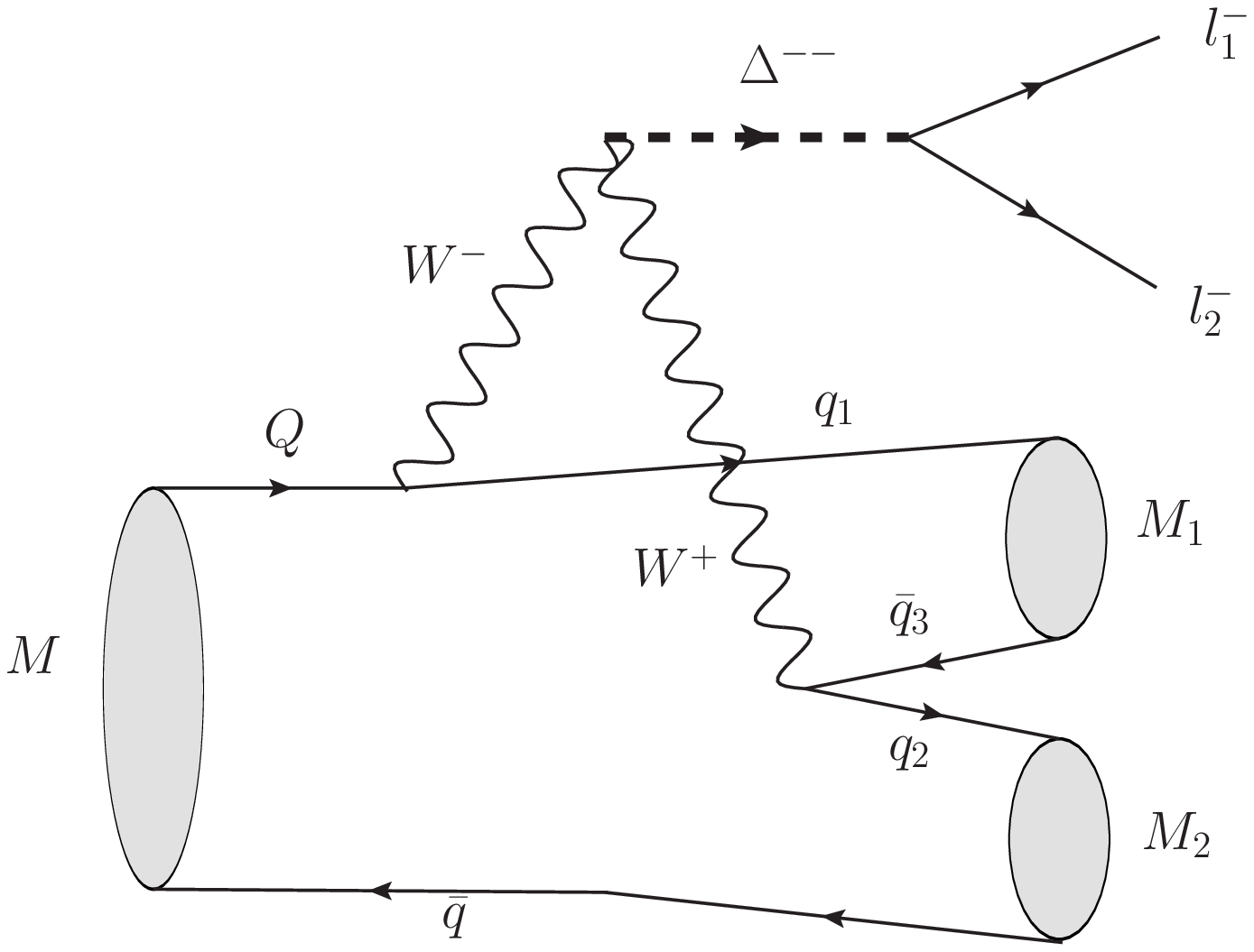}}
\caption[]{Feynman diagrams of the decay processes $h\rightarrow h_1h_2l_1^-l_2^-$.}
\end{figure}

These channels can also be induced by Majorana-type neutrinos. Their Feynman diagrams are similar to Fig.~1, but the $s$ channels should be replaced by $t$ channels. If the neutrino mass is around GeV, it could be produced on-shell, which has attracted many attentions~\cite{yuan13, yuan17, dong15, cas13}. For the cases when the mass is very small or very large, the branching ratios will have the same order of magnitude as that of the doubly-charged Higgs boson case~\cite{pic97,ali01}.  So theoretical consideration of these low energy processes induced by the doubly-charged Higgs boson will provide a useful supplement to the Majorana neutrino scenario.

This work is organized as follows. In Sec. II, we give the Lagrangian which describes the coupling between the Higgs triplet and the standard model particles. Then the amplitudes and phase space integral of four-body decays of the neural flavored mesons are presented. In Sec. III, we give the branching ratios of all the decay channels and compare the results of $D^0$ with the experimental data. The last section is reserved for the conclusion.  Some details for the wave functions of mesons are presented in the Appendix. 

\section{Theoretical Formalism}

 The assumed Higgs triplet $\Delta$ in the $2\times 2$ representation is defined as~\cite{ma09}
\begin{equation}
\Delta = \left(
\begin{array}{cc}
\Delta^+/\sqrt{2}& \Delta^{++} \\
\Delta^0 & -\Delta^+/\sqrt{2}\\
\end{array}
\right).
\end{equation}
It mixes with the usual $SU(2)_L$ Higgs doublet by a mixing angle $\theta_\Delta$, from which we define $s_\Delta=\sin\theta_\Delta$ and $c_\Delta=\cos\theta_\Delta$.

The Lagrangian which describes the interaction between $\Delta$ and $W^-$ gauge boson or SM fermions has the following form~\cite{pic97, ma09}
\begin{equation}
\begin{aligned}
\mathcal L_{int}^\prime&=ih_{ij}\psi_{iL}^TC\sigma_2\Delta\psi_{jL}-\sqrt{2}gm_Ws_\Delta \Delta^{++}W^{-\mu}W^-_\mu+ \frac{\sqrt{2}}{2}gc_\Delta W^{-\mu}\Delta^-\overset{\leftrightarrow}{\partial}_\mu\Delta^{++}\\
&~~~~+\frac{igs_\Delta}{\sqrt{2}m_Wc_\Delta}\Delta^+(m_{q^\prime}\bar q_Rq_R^\prime-m_q\bar q_Lq_L^\prime)+ H.c.,
\end{aligned}
\end{equation}
where $C=i\gamma^2\gamma^0$ is the charge conjugation matrix; $\psi_{iL}$ represents the leptonic doublet; $h_{ij}$ is the leptonic Yukawa coupling constant; $g$ is the weak coupling constant. Compared with the second term, the third and the fourth terms can be neglected as they give smaller contributions.

If $q=q_3$, all the four diagrams of Fig. 1 will contribute:
\begin{equation}
\begin{aligned}
\mathcal M_A&= \frac{g^3}{8\sqrt{2}m_W^3}V_{q_1Q}V_{q_2q_3} \frac{s_\Delta h_{ij}}{m_{\Delta}^2}\langle h_1(p_1)h_2(p_2)| (\bar q_1Q)_{_{V-A}} (\bar q_2q_3)_{_{V-A}} |h(p)\rangle \langle lepton\rangle\\
&=\frac{g^3}{8\sqrt{2}m_W^3}V_{q_1Q}V_{q_2q_3} \frac{s_\Delta h_{ij}}{m_{\Delta}^2}f_{h_2} p_2^\mu\langle h_1(p_1)| \bar q_1\gamma_\mu(1-\gamma_5)Q |h(p)\rangle \langle lepton\rangle,
\end{aligned}
\end{equation}
\begin{equation}
\begin{aligned}
\mathcal M_B&= \frac{g^3}{8\sqrt{2}m_W^3}V_{q_2Q}V_{q_1q_3} \frac{s_\Delta h_{ij}}{m_{\Delta}^2}\langle h_1(p_1)h_2(p_2)| (\bar q_2Q)_{_{V-A}} (\bar q_1q_3)_{_{V-A}} | h(p)\rangle \langle lepton\rangle\\
&=\frac{g^3}{8\sqrt{2}m_W^3}\frac{1}{3}V_{q_2Q}V_{q_1q_3} \frac{s_\Delta h_{ij}}{m_{\Delta}^2}f_{h_2} p_2^\mu\langle h_1(p_1)| \bar q_1\gamma_\mu(1-\gamma_5)Q |h(p)\rangle \langle lepton\rangle,
\end{aligned}
\end{equation}
\begin{equation}
\begin{aligned}
\mathcal M_C&= \frac{g^3}{8\sqrt{2}m_W^3}V_{q_2Q}V_{q_1q_3} \frac{s_\Delta h_{ij}}{m_{\Delta}^2}\langle h_1(p_1)h_2(p_2)| (\bar q_2Q)_{_{V-A}} (\bar q_1q_3)_{_{V-A}} | h(p)\rangle \langle lepton\rangle\\
&=\frac{g^3}{8\sqrt{2}m_W^3}V_{q_2Q}V_{q_1q_3} \frac{s_\Delta h_{ij}}{m_{\Delta}^2}f_{h_1} p_1^\mu\langle h_2(p_2)| \bar q_2\gamma_\mu(1-\gamma_5)Q | h(p)\rangle \langle lepton\rangle,
\end{aligned}
\end{equation}
\begin{equation}
\begin{aligned}
\mathcal M_D&= \frac{g^3}{8\sqrt{2}m_W^3}V_{q_1Q}V_{q_2q_3} \frac{s_\Delta h_{ij}}{m_{\Delta}^2}\langle h_1(p_1)h_2(p_2)| (\bar q_1q_3)_{_{V-A}} (\bar q_2Q)_{_{V-A}} | h(p)\rangle \langle lepton\rangle\\
&=\frac{g^3}{8\sqrt{2}m_W^3}\frac{1}{3}V_{q_1Q}V_{q_2q_3} \frac{s_\Delta h_{ij}}{m_{\Delta}^2}f_{h_1} p_1^\mu\langle h_2(p_2)| \bar q_2\gamma_\mu(1-\gamma_5)Q | h(p)\rangle \langle lepton\rangle,
\end{aligned}
\end{equation}
where the factor $\frac{1}{3}$ in $\mathcal M_B$ and $\mathcal M_D$ is introduced by the Fierz transformation; $\langle lepton\rangle$ is the leptonic part of the transition matrix element; $V_{q_iq_j}$ is the Cabibbo-Kobayashi-Maskawa matrix element. The definition of the decay constant $f_{h_1}$ of a pseudoscalar meson 
\begin{equation}
\begin{aligned}
\langle h_1(p_1)| \bar q_1\gamma^\mu(1-\gamma_5) q_2|0\rangle = if_{h_1} p_1^\mu
\end{aligned}
\end{equation}
is used. For vector mesons, it should be replaced by 
\begin{equation}
\begin{aligned}
\langle h_1(p_1, \epsilon)| \bar q_1\gamma^\mu(1-\gamma_5) q_2|0\rangle = M_1f_{h_1} \epsilon^\mu.
\end{aligned}
\end{equation} 
In Table I, the values of the decay constants are presented.  
\begin{table}
\caption{Decay constants (in units of MeV) of mesons. Those for $\pi$, $K$, $D$, and $D_s$ are from Particle Data Group~\cite{pdg}; $K^\ast$ and $\rho$,  are from Ref.~\cite{ball}; $D^\ast$ and $D_s^\ast$ are from Ref.~\cite{wang06}.} 
\label{results}
\setlength{\tabcolsep}{0.1cm}
\centering
\begin{tabular*}{\textwidth}{@{}@{\extracolsep{\fill}}cccccccc}
\hline\hline
{\phantom{\Large{l}}}\raisebox{+.2cm}{\phantom{\Large{j}}}
$f_\pi$&$f_K$&$f_{K^\ast}$&$f_\rho$&$f_D$&$f_{D_s}$&$f_{D^\ast}$&$f_{D_s^\ast}$\\ 
{\phantom{\Large{l}}}\raisebox{+.3cm}{\phantom{\Large{j}}}
$130.4$&$156.2$&$217$&$205$&$204.6$&$257.5$&$340$&$375$\\ 
\hline\hline
\end{tabular*}
\end{table}
Finally, we get the transition amplitude
\begin{equation}
\begin{aligned}
\mathcal M&=\mathcal M_A+\mathcal M_B+\mathcal M_C+\mathcal M_D\\
&=\frac{g^3s_\Delta h_{ij}}{8\sqrt{2}m_W^3m_{\Delta}^2}\Big\{(V_{q_1Q}V_{q_2q_3}+\frac{1}{3}V_{q_2Q}V_{q_1q_3})f_{h_2} p_2^\mu\langle h_1(p_1)| \bar q_1\gamma_\mu(1-\gamma_5)Q |h(p)\rangle\\
&+(V_{q_2Q}V_{q_1q_3}+\frac{1}{3}V_{q_1Q}V_{q_2q_3})f_{h_1} p_1^\mu\langle h_2(p_2)| \bar q_2\gamma_\mu(1-\gamma_5)Q |h(p)\rangle\Big\}\langle lepton\rangle.
\end{aligned}
\end{equation}
If $q\ne q_3$, only Fig. 1(A) and (B) contribute: 
\begin{equation}
\begin{aligned}
\mathcal M&=\mathcal M_A+\mathcal M_B\\
&=\frac{g^3s_\Delta h_{ij}}{8\sqrt{2}m_W^3m_{\Delta}^2}(V_{q_1Q}V_{q_2q_3}+\frac{1}{3}V_{q_2Q}V_{q_1q_3})f_{h_2} p_2^\mu\langle h_1(p_1)| \bar q_1\gamma_\mu(1-\gamma_5)Q | h(p)\rangle\langle lepton\rangle.
\end{aligned}
\end{equation}

The hadronic transition matrix can be expressed as~\cite{fu12}
\begin{equation}
\begin{aligned}
\langle h_1(p_1)| V^\mu |h(p)\rangle&=  f_+(Q^2)(p+p_1)^\mu+f_-(Q^2)(p-p_1)^\mu,
\end{aligned}
\end{equation}
for $h_1$ being a pseudoscalar meson, where $f_+$ and $f_-$ are form factors.
If $h_1$ is a vector meson, we have
\begin{equation}
\begin{aligned}
&\langle h_1(p_1,\epsilon)| V^\mu |h(p)\rangle= -i\frac{2}{M+M_1}f_V(Q^2)\epsilon^{\mu\epsilon^\ast pp_1},\\
&\langle h_1(p_1,\epsilon)| A^\mu |h(p)\rangle= f_1(Q^2)\frac{\epsilon^\ast\cdot p}{M+M_1}(p+p_1)^\mu+f_2(Q^2)\frac{\epsilon^\ast\cdot p}{M+M_1}(p-p_1)^\mu\\
&~~~~~~~~~~~~~~~~~~~~~~~~~~~+ f_0(Q^2)(M+M_1)\epsilon^{\ast\mu},
\end{aligned}
\end{equation}
where $f_V$ and $f_i$ ($i=0,~1,~2$) are form factors; $M$ and $M_1$ are the masses of corresponding mesons; the definition $Q=p-p_1$ is used.

By applying the Bethe-Salpeter method with the instantaneous approximation~\cite{chang01}, the hadronic matrix element is written as 
\begin{equation}
\begin{aligned}
\langle h_1(p_1)|\bar q_1\gamma^\mu(1-\gamma_5)Q|h(p)\rangle=\int\frac{d^3q}{(2\pi)^{3}}\textrm{Tr}\left[\frac{\slashed p}{M}\overline{\varphi_{p_1}^{++}}({\vec{q}_1})\gamma_{\mu}(1-\gamma_{5})\varphi_{p}^{++}({\vec{q}})\right],
\end{aligned}
\end{equation}
where $\varphi^{++}$ is the positive energy part of the wave function whose expression can be found in the Appendix; $\vec q$ and $\vec q_1$ are the relative three-momenta between the quark and antiquark in the initial and final mesons, respectively.

The partial decay width is achieved by finishing the phase space integral
\begin{equation}
\begin{aligned}
\Gamma = (1-\frac{1}{2}\delta_{h_1h_2})(1-\frac{1}{2}\delta_{l_1l_2})\int \frac{ds_{12}}{s_{12}}\int\frac{ds_{34}}{s_{34}} \int d\cos\theta_{12} \int d\cos\theta_{34} \int d\phi  \mathcal K|\mathcal M|^2,
\end{aligned}
\end{equation}
where 
\begin{equation}
\mathcal K = \frac{1}{2^{15}\pi^6 M^3} \lambda^{1/2}(M^2, s_{12}, s_{34}) \lambda^{1/2}(s_{12}, M_1^2, M_2^2)\lambda^{1/2}(s_{34},m_1^2, m_2^2).
\end{equation}
We also use the definitions $s_{12}= (p_1+p_2)^2$ and $s_{34}= (p_3+p_4)^2$. The meanings of $\theta_{12}$, $\theta_{34}$, and $\phi$ are explicit from Fig. 2. $\delta_{l_1l_2}$ is 1 if $l_1$ and $l_2$ are identical particles, otherwise, it is 0. The same is true for $\delta_{h_1h_2}$.
The integral limits are
\begin{equation}
\begin{aligned}
&s_{12}\in[(M_1+M_2)^2,~(M-m_1-m_2)^2],\\
&s_{34}\in[(m_1+m_2)^2,~(M-\sqrt{s_{12}})^2],\\
&\phi\in[0,~2\pi],~~~\theta_{12}\in[0,~\pi],~~~\theta_{34}\in[0,~\pi],
\end{aligned}
\end{equation}
where $M_2$, $m_1$, and $m_2$ are the masses of $h_2$, $l_1$, and $l_2$, respectively. 

\begin{figure}[ht]
\centering
\includegraphics[scale=0.8]{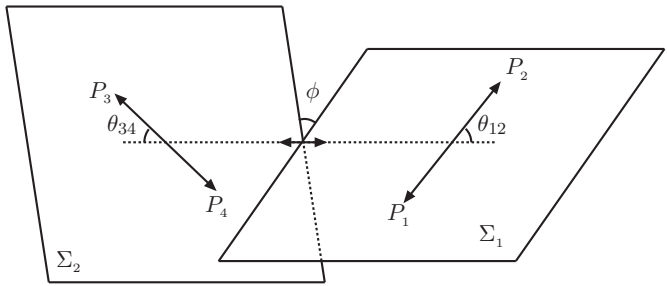}
\caption[]{Kinematics of the four-body decay of $h$ in its rest frame. $P_1$ and $P_2$ are respectively the momenta of $h_1$ and $h_2$ in their center-of-momentum frame; $P_3$ and $P_4$ are respectively the momenta of $l_1$ and $l_2$ in their center-of-momentum frame.}
\end{figure}

\section{Numerical Results}

The wave functions of the initial and final mesons in the hadronic transition matrix element are evaluated numerically by solving corresponding instantaneous Bethe-Salpeter equation. The interaction kernel we adopt is a Cornell-type potential whose expression is presented in the Appendix. Strictly speaking, although the instantaneous approximation is reasonable for the double heavy mesons and acceptable for the heavy-light mesons, it will bring large errors for the light mesons, such as $\pi$ and $K$. Nevertheless, we also apply this approximation to the light mesons to make the estimation, as the decay channels considered here are related to the new physics, of which only the order of magnitude is important. This is also the reason why we do not consider the QCD corrections and the final meson interactions as mentioned in Ref.~\cite{wang18}.

The related parameters of the doubly-charged Higgs boson have no definite values until now. Only the lower or upper limits from experiments exist. For example, the latest results of the ALTAS and CMS Collaborations~\cite{ATLAS, CMS} show that the mass of $\Delta^{++}$ is more than 800 GeV. If we set its mass to $1000$ GeV, and adopt the same values of $s_\Delta$ and $h_{ij}$ in Ref.~\cite{wang18}, we can estimate $\left(\frac{s_\Delta h_{ij}}{m_\Delta^2}\right)^{2}$ to be less than $10^{-16}$ for $ee$ and $\mu\mu$, and $10^{-26}$ for $e\mu$. Here we present the results for the quantity $Br\times\left(\frac{s_\Delta h_{ij}}{m_\Delta^2}\right)^{-2}$, where $Br$ represents the branching ratio whose upper limit can be easily achieved by using the parameter mentioned above. 

For $\bar K^0$, there are only three such channels allowed by the phase space, namely $\pi^+\pi^+l_1^-l_2^-~(l_i=e,~\mu)$ . The corresponding diagrams are Fig.~1(A)$\sim$(D). $Br\times\left(\frac{s_\Delta h_{ij}}{m_\Delta^2}\right)^{-2}$ is of the order of  $10^{-11}$ ${\rm GeV^{4}}$ (see Table II). Experimentally, $Br(K^+\rightarrow \pi^-l_1^+l_2^+)\lesssim 10^{-10}$~\cite{app}, which is the most precise result of the lepton number violation decay channels. However, there is no experimental detections of lepton number violation four-body decay channels of this particle. In Refs.~\cite{NA48, KTeV}, the channels $K_{L,S}\rightarrow \pi^+\pi^-e^+e^-$ are investigated. We expect there will be experiments on $K_{L,S}\rightarrow \pi^+\pi^+l_1^-l_2^-$ channels. 

For $D^0$, the final mesons can be pseudoscalars or vectors. When $h_1$ and $h_2$ are both pseudoscalars, that is $\pi\pi$, $\pi K$, or $KK$, the results are given in Table III. The largest value has the order of magnitude $10^{-10}$ ${\rm GeV^4}$. One notices that the Fermilab E791 Collaboration once presented the upper limits of the branching ratios of these channels~\cite{E791}, which are of the order of $10^{-5}$. By comparing the theoretical prediction and the experimental data,  we can set the upper limit of the constant $\frac{s_\Delta h_{ij}}{m_\Delta^2}$, which is of the order of $10^3$ ${\rm GeV^{-2}}$.  One can also extract this upper limit from the three-body decay processes, such as $D^-\to\pi^+e^-e^-$, which is about $10^2$ ${\rm GeV^{-2}}$ by using the results in Ref.~\cite{ma09}. The decay channels of $D^0$ with $h_1$ and $h_2$ being $0^-1^-$ or $1^-1^-$ are presented in Table  IV, which have the largest value about $10^{-11}$ ${\rm GeV^4}$.

The results for $\bar B^0$ and $\bar B_s^0$ are presented in Table V$\sim$X. The largest value is also of the order of $10^{-10}$ ${\rm GeV^4}$. In Ref.~\cite{LHCb}, the four-body decay channel $B^-\rightarrow D^0\pi^+\mu^-\mu^-$ are measured to have the branching ratio less than $1.5\times 10^{-6}$. There is no experimental results for the neutral $B$ meson decay channels until now. However, as LHCb running, more and more data will be accumulated. We expect that the LHCb Collaboration will detect such decay modes and set more stringent constraint on the parameters of doubly-charged Higgs boson. Besides that, the future B-factories, such as Belle-II, also has the possibility to provide more information about such channels.

\begin{table}
\caption{$Br\times\left(\frac{s_\Delta h_{ij}}{m_\Delta^2}\right)^{-2}$ for different decay channels of $\bar K^0$.}
\vspace{0.2cm}
\setlength{\tabcolsep}{0.01cm}
\centering
\begin{tabular*}{\textwidth}{@{}@{\extracolsep{\fill}}cc}
\hline\hline
decay channel&$Br\times\left(\frac{s_\Delta h_{ij}}{m_\Delta^2}\right)^{-2}$ (${\rm GeV^4}$)\\
\hline
{\phantom{\Large{l}}}\raisebox{+.2cm}{\phantom{\Large{j}}}
$\bar K^0\rightarrow \pi^+\pi^+e^-e^-$&$2.23\times10^{-11}$\\
{\phantom{\Large{l}}}\raisebox{+.2cm}{\phantom{\Large{j}}}
$\bar K^0\rightarrow \pi^+\pi^+\mu^-\mu^-$&$1.58\times10^{-13}$\\
{\phantom{\Large{l}}}\raisebox{+.2cm}{\phantom{\Large{j}}}
$\bar K^0\rightarrow \pi^+\pi^+e^-\mu^-$&$1.15\times10^{-11}$\\
\hline\hline
\end{tabular*}
\end{table}

\begin{table}
\caption{$Br\times\left(\frac{s_\Delta h_{ij}}{m_\Delta^2}\right)^{-2}$ for $0^-0^-$ decay channels of $D^0$.}
\vspace{0.2cm}
\setlength{\tabcolsep}{0.01cm}
\centering
\begin{tabular*}{\textwidth}{@{}@{\extracolsep{\fill}}cccc}
\hline\hline
decay channel&$Br\times\left(\frac{s_\Delta h_{ij}}{m_\Delta^2}\right)^{-2}$ (${\rm GeV^4}$) &Exp. bound on $Br$~\cite{E791} &$\frac{s_\Delta h_{ij}}{m_\Delta^2}$ (${\rm GeV^{-2}}$)\\ \hline
{\phantom{\Large{l}}}\raisebox{+.2cm}{\phantom{\Large{j}}}
$D^0\rightarrow \pi^-\pi^-e^+e^+$&$1.79\times 10^{-11}$ &$<11.2\times 10^{-5}$&$<2501$\\
{\phantom{\Large{l}}}\raisebox{+.2cm}{\phantom{\Large{j}}}
$D^0\rightarrow \pi^-\pi^-\mu^+\mu^+$&$1.93\times 10^{-11}$ &$<2.9\times 10^{-5}$&$<1227$\\
{\phantom{\Large{l}}}\raisebox{+.2cm}{\phantom{\Large{j}}}
$D^0\rightarrow \pi^-\pi^-e^+\mu^+$& $3.61\times 10^{-11}$ &$<7.9\times 10^{-5}$&$<1479$\\
{\phantom{\Large{l}}}\raisebox{+.2cm}{\phantom{\Large{j}}}
$D^0\rightarrow \pi^-K^-e^+e^+$& $9.96\times 10^{-11}$ &$<20.6\times 10^{-5}$&$<1438$\\
{\phantom{\Large{l}}}\raisebox{+.2cm}{\phantom{\Large{j}}}
$D^0\rightarrow \pi^-K^-\mu^+\mu^+$& $1.08\times 10^{-10}$ &$<39.0\times 10^{-5}$&$<1898$\\
{\phantom{\Large{l}}}\raisebox{+.2cm}{\phantom{\Large{j}}}
$D^0\rightarrow \pi^-K^-e^+\mu^+$& $2.01\times 10^{-10}$ &$<21.8\times 10^{-5}$&$<1041$\\
{\phantom{\Large{l}}}\raisebox{+.2cm}{\phantom{\Large{j}}}
$D^0\rightarrow K^-K^-e^+e^+$& $1.02\times 10^{-11}$ &$<15.2\times 10^{-5}$&$<3869$\\
{\phantom{\Large{l}}}\raisebox{+.2cm}{\phantom{\Large{j}}}
$D^0\rightarrow K^-K^-\mu^+\mu^+$& $1.11\times 10^{-11}$ &$<9.4\times 10^{-5}$&$<2914$\\
{\phantom{\Large{l}}}\raisebox{+.2cm}{\phantom{\Large{j}}}
$D^0\rightarrow K^-K^-e^+\mu^+$& $2.03\times 10^{-11}$ &$<5.7\times 10^{-5}$&$<1677$\\
\hline\hline
\end{tabular*}
\end{table}

\begin{table}
\caption{$Br\times\left(\frac{s_\Delta h_{ij}}{m_\Delta^2}\right)^{-2}$ for $0^-1^-$ and $1^-1^-$ decay channels of $D^0$.}
\vspace{0.2cm}
\setlength{\tabcolsep}{0.01cm}
\centering
\begin{tabular*}{\textwidth}{@{}@{\extracolsep{\fill}}cccc}
\hline\hline
decay channel&$Br\times\left(\frac{s_\Delta h_{ij}}{m_\Delta^2}\right)^{-2}$ (${\rm GeV^4}$)&decay channel& $Br\times\left(\frac{s_\Delta h_{ij}}{m_\Delta^2}\right)^{-2}$ (${\rm GeV^4}$)\\ \hline
{\phantom{\Large{l}}}\raisebox{+.2cm}{\phantom{\Large{j}}}
$D^0\rightarrow \pi^-\rho^-e^+e^+$&$3.37\times 10^{-12}$&$D^0\rightarrow \rho^-\rho^-e^+e^+$&$2.24\times 10^{-15}$\\
{\phantom{\Large{l}}}\raisebox{+.2cm}{\phantom{\Large{j}}}
$D^0\rightarrow \pi^-\rho^-\mu^+\mu^+$&$3.15\times 10^{-12}$&$D^0\rightarrow \rho^-\rho^-\mu^+\mu^+$&$1.16\times 10^{-15}$\\
{\phantom{\Large{l}}}\raisebox{+.2cm}{\phantom{\Large{j}}}
$D^0\rightarrow \pi^-\rho^-e^+\mu^+$&$6.17\times 10^{-12}$&$D^0\rightarrow \rho^-\rho^-e^+\mu^+$&$2.85\times 10^{-15}$\\
{\phantom{\Large{l}}}\raisebox{+.2cm}{\phantom{\Large{j}}}
$D^0\rightarrow \pi^-K^{\ast-}e^+e^+$&$2.12\times 10^{-13}$&$D^0\rightarrow \rho^-K^{\ast-}e^+e^+$&$6.76\times 10^{-15}$\\
{\phantom{\Large{l}}}\raisebox{+.2cm}{\phantom{\Large{j}}}
$D^0\rightarrow \pi^-K^{\ast-}\mu^+\mu^+$&$1.86\times 10^{-13}$&$D^0\rightarrow \rho^-K^{\ast-}e^+\mu^+$&$3.71\times 10^{-15}$\\
{\phantom{\Large{l}}}\raisebox{+.2cm}{\phantom{\Large{j}}}
$D^0\rightarrow \pi^-K^{\ast-}e^+\mu^+$&$3.72\times 10^{-13}$&$D^0\rightarrow K^-K^{\ast-}e^+e^+$&$5.40\times 10^{-13}$\\
{\phantom{\Large{l}}}\raisebox{+.2cm}{\phantom{\Large{j}}}
$D^0\rightarrow \rho^-K^{-}e^+e^+$&$2.08\times 10^{-11}$&$D^0\rightarrow K^-K^{\ast-}\mu^+\mu^+$&$3.68\times 10^{-13}$\\
{\phantom{\Large{l}}}\raisebox{+.2cm}{\phantom{\Large{j}}}
$D^0\rightarrow \rho^-K^{-}\mu^+\mu^+$&$1.76\times 10^{-11}$&$D^0\rightarrow K^-K^{\ast-}e^+\mu^+$&$8.03\times 10^{-13}$\\
{\phantom{\Large{l}}}\raisebox{+.2cm}{\phantom{\Large{j}}}
$D^0\rightarrow \rho^-K^{-}e^+\mu^+$&$3.52\times 10^{-11}$&$D^0\rightarrow K^{\ast-}K^{\ast-}e^+e^+$&$4.13\times 10^{-17}$\\
\hline\hline
\end{tabular*}
\end{table}

\begin{table}
\caption{$Br\times\left(\frac{s_\Delta h_{ij}}{m_\Delta^2}\right)^{-2}$ for $0^-0^-$ decay channels of $\bar B^0$.}
\vspace{0.2cm}
\setlength{\tabcolsep}{0.01cm}
\centering
\begin{tabular*}{\textwidth}{@{}@{\extracolsep{\fill}}cccc}
\hline\hline
decay channel&$Br\times\left(\frac{s_\Delta h_{ij}}{m_\Delta^2}\right)^{-2}$ (${\rm GeV^4}$)&decay channel& $Br\times\left(\frac{s_\Delta h_{ij}}{m_\Delta^2}\right)^{-2}$ (${\rm GeV^4}$)\\ \hline
{\phantom{\Large{l}}}\raisebox{+.2cm}{\phantom{\Large{j}}}
$\bar B^0\rightarrow \pi^+\pi^+e^-e^-$&$7.62\times 10^{-13}$&$\bar B^0\rightarrow \pi^+D_s^+e^-\mu^-$&$3.43\times 10^{-12}$\\
{\phantom{\Large{l}}}\raisebox{+.2cm}{\phantom{\Large{j}}}
$\bar B^0\rightarrow \pi^+\pi^+\mu^-\mu^-$&$8.18\times 10^{-13}$&$\bar B^0\rightarrow K^+D^+e^-e^-$&$1.02\times 10^{-12}$\\
{\phantom{\Large{l}}}\raisebox{+.2cm}{\phantom{\Large{j}}}
$\bar B^0\rightarrow \pi^+\pi^+e^-\mu^-$&$1.58\times 10^{-12}$&$\bar B^0\rightarrow K^+D^+\mu^-\mu^-$&$1.11\times 10^{-12}$\\
{\phantom{\Large{l}}}\raisebox{+.2cm}{\phantom{\Large{j}}}
$\bar B^0\rightarrow \pi^+K^+e^-e^-$&$3.74\times 10^{-14}$&$\bar B^0\rightarrow K^+D^+e^-\mu^-$&$2.11\times 10^{-12}$\\
{\phantom{\Large{l}}}\raisebox{+.2cm}{\phantom{\Large{j}}}
$\bar B^0\rightarrow \pi^+K^+\mu^-\mu^-$&$4.02\times 10^{-14}$&$\bar B^0\rightarrow D^+D^+e^-e^-$&$2.27\times 10^{-12}$\\
{\phantom{\Large{l}}}\raisebox{+.2cm}{\phantom{\Large{j}}}
$\bar B^0\rightarrow \pi^+K^+e^-\mu^-$&$7.74\times 10^{-14}$&$\bar B^0\rightarrow D^+D^+\mu^-\mu^-$&$2.58\times 10^{-12}$\\
{\phantom{\Large{l}}}\raisebox{+.2cm}{\phantom{\Large{j}}}
$\bar B^0\rightarrow \pi^+D^+e^-e^-$&$1.35\times 10^{-11}$&$\bar B^0\rightarrow D^+D^+e^-\mu^-$&$4.79\times 10^{-12}$\\
{\phantom{\Large{l}}}\raisebox{+.2cm}{\phantom{\Large{j}}}
$\bar B^0\rightarrow \pi^+D^+\mu^-\mu^-$&$1.46\times 10^{-11}$&$\bar B^0\rightarrow D^+D_s^+e^-e^-$&$3.68\times 10^{-11}$\\
{\phantom{\Large{l}}}\raisebox{+.2cm}{\phantom{\Large{j}}}
$\bar B^0\rightarrow \pi^+D^+e^-\mu^-$&$2.80\times 10^{-11}$&$\bar B^0\rightarrow D^+D_s^+\mu^-\mu^-$&$4.20\times 10^{-11}$\\
{\phantom{\Large{l}}}\raisebox{+.2cm}{\phantom{\Large{j}}}
$\bar B^0\rightarrow \pi^+D_s^+e^-e^-$&$1.65\times 10^{-12}$&$\bar B^0\rightarrow D^+D_s^+e^-\mu^-$&$7.76\times 10^{-11}$\\
{\phantom{\Large{l}}}\raisebox{+.2cm}{\phantom{\Large{j}}}
$\bar B^0\rightarrow \pi^+D_s^+\mu^-\mu^-$&$1.79\times 10^{-12}$&&\\
\hline\hline
\end{tabular*}
\end{table}

\begin{table}
\caption{$Br\times\left(\frac{s_\Delta h_{ij}}{m_\Delta^2}\right)^{-2}$ for $0^-1^-$ decay channels of $\bar B^0$.}
\vspace{0.2cm}
\setlength{\tabcolsep}{0.01cm}
\centering
\begin{tabular*}{\textwidth}{@{}@{\extracolsep{\fill}}cccc}
\hline\hline
decay channel&$Br\times\left(\frac{s_\Delta h_{ij}}{m_\Delta^2}\right)^{-2}$ (${\rm GeV^4}$)&decay channel& $Br\times\left(\frac{s_\Delta h_{ij}}{m_\Delta^2}\right)^{-2}$ (${\rm GeV^4}$)\\ \hline
{\phantom{\Large{l}}}\raisebox{+.2cm}{\phantom{\Large{j}}}
$\bar B^0\rightarrow \pi^+\rho^+e^-e^-$&$9.53\times 10^{-13}$&$\bar B^0\rightarrow \rho^+D_s^{+}e^-e^-$&$1.27\times 10^{-15}$\\
{\phantom{\Large{l}}}\raisebox{+.2cm}{\phantom{\Large{j}}}
$\bar B^0\rightarrow \pi^+\rho^+\mu^-\mu^-$&$1.02\times 10^{-12}$&$\bar B^0\rightarrow \rho^+D_s^{+}\mu^-\mu^-$&$1.37\times 10^{-15}$\\
{\phantom{\Large{l}}}\raisebox{+.2cm}{\phantom{\Large{j}}}
$\bar B^0\rightarrow \pi^+\rho^+e^-\mu^-$&$1.97\times 10^{-12}$&$\bar B^0\rightarrow \rho^+D_s^{+}e^-\mu^-$&$2.63\times 10^{-15}$\\
{\phantom{\Large{l}}}\raisebox{+.2cm}{\phantom{\Large{j}}}
$\bar B^0\rightarrow \pi^+K^{\ast+}e^-e^-$&$5.58\times 10^{-14}$&$\bar B^0\rightarrow K^+D^{\ast+}e^-e^-$&$1.33\times 10^{-14}$\\
{\phantom{\Large{l}}}\raisebox{+.2cm}{\phantom{\Large{j}}}
$\bar B^0\rightarrow \pi^+K^{\ast+}\mu^-\mu^-$&$5.99\times 10^{-14}$&$\bar B^0\rightarrow K^+D^{\ast+}\mu^-\mu^-$&$1.41\times 10^{-14}$\\
{\phantom{\Large{l}}}\raisebox{+.2cm}{\phantom{\Large{j}}}
$\bar B^0\rightarrow \pi^+K^{\ast+}e^-\mu^-$&$1.15\times 10^{-13}$&$\bar B^0\rightarrow K^+D^{\ast+}e^-\mu^-$&$2.72\times 10^{-14}$\\
{\phantom{\Large{l}}}\raisebox{+.2cm}{\phantom{\Large{j}}}
$\bar B^0\rightarrow \rho^+K^{+}e^-e^-$&$6.09\times 10^{-17}$&$\bar B^0\rightarrow K^{\ast+}D^{+}e^-e^-$&$1.27\times 10^{-12}$\\
{\phantom{\Large{l}}}\raisebox{+.2cm}{\phantom{\Large{j}}}
$\bar B^0\rightarrow \rho^+K^{+}\mu^-\mu^-$&$6.52\times 10^{-17}$&$\bar B^0\rightarrow K^{\ast+}D^{+}\mu^-\mu^-$&$1.39\times 10^{-12}$\\
{\phantom{\Large{l}}}\raisebox{+.2cm}{\phantom{\Large{j}}}
$\bar B^0\rightarrow \rho^+K^{+}e^-\mu^-$&$1.26\times 10^{-16}$&$\bar B^0\rightarrow K^{\ast+}D^{+}e^-\mu^-$&$2.64\times 10^{-12}$\\
{\phantom{\Large{l}}}\raisebox{+.2cm}{\phantom{\Large{j}}}
$\bar B^0\rightarrow \pi^+D^{\ast+}e^-e^-$&$9.44\times 10^{-14}$&$\bar B^0\rightarrow D^+D^{\ast+}e^-e^-$&$5.56\times 10^{-13}$\\
{\phantom{\Large{l}}}\raisebox{+.2cm}{\phantom{\Large{j}}}
$\bar B^0\rightarrow \pi^+D^{\ast+}\mu^-\mu^-$&$9.87\times 10^{-14}$&$\bar B^0\rightarrow D^+D^{\ast+}\mu^-\mu^-$&$6.24\times 10^{-13}$\\
{\phantom{\Large{l}}}\raisebox{+.2cm}{\phantom{\Large{j}}}
$\bar B^0\rightarrow \pi^+D^{\ast+}e^-\mu^-$&$1.91\times 10^{-13}$&$\bar B^0\rightarrow D^+D^{\ast+}e^-\mu^-$&$1.16\times 10^{-12}$\\
{\phantom{\Large{l}}}\raisebox{+.2cm}{\phantom{\Large{j}}}
$\bar B^0\rightarrow \rho^+D^{+}e^-e^-$&$2.32\times 10^{-11}$&$\bar B^0\rightarrow D^+D_s^{\ast+}e^-e^-$&$1.22\times 10^{-11}$\\
{\phantom{\Large{l}}}\raisebox{+.2cm}{\phantom{\Large{j}}}
$\bar B^0\rightarrow \rho^+D^{+}\mu^-\mu^-$&$2.53\times 10^{-11}$&$\bar B^0\rightarrow D^+D_s^{\ast+}\mu^-\mu^-$&$1.37\times 10^{-11}$\\
{\phantom{\Large{l}}}\raisebox{+.2cm}{\phantom{\Large{j}}}
$\bar B^0\rightarrow \rho^+D^{+}e^-\mu^-$&$4.82\times 10^{-11}$&$\bar B^0\rightarrow D^+D_s^{\ast+}e^-\mu^-$&$2.55\times 10^{-11}$\\
{\phantom{\Large{l}}}\raisebox{+.2cm}{\phantom{\Large{j}}}
$\bar B^0\rightarrow \pi^+D_s^{\ast+}e^-e^-$&$1.17\times 10^{-12}$&$\bar B^0\rightarrow D^{\ast+}D_s^{+}e^-e^-$&$6.39\times10^{-14}$\\
{\phantom{\Large{l}}}\raisebox{+.2cm}{\phantom{\Large{j}}}
$\bar B^0\rightarrow \pi^+D_s^{\ast+}\mu^-\mu^-$&$1.26\times 10^{-12}$&$\bar B^0\rightarrow D^{\ast+}D_s^{+}\mu^-\mu^-$&$6.64\times 10^{-14}$\\
{\phantom{\Large{l}}}\raisebox{+.2cm}{\phantom{\Large{j}}}
$\bar B^0\rightarrow \pi^+D_s^{\ast+}e^-\mu^-$&$2.42\times 10^{-12}$&$\bar B^0\rightarrow D^{\ast+}D_s^{+}e^-\mu^-$&$1.27\times 10^{-13}$\\
\hline\hline
\end{tabular*}
\end{table}

\begin{table}
\caption{$Br\times\left(\frac{s_\Delta h_{ij}}{m_\Delta^2}\right)^{-2}$ for $1^-1^-$ decay channels of $\bar B^0$.}
\vspace{0.2cm}
\setlength{\tabcolsep}{0.01cm}
\centering
\begin{tabular*}{\textwidth}{@{}@{\extracolsep{\fill}}cccc}
\hline\hline
decay channel&$Br\times\left(\frac{s_\Delta h_{ij}}{m_\Delta^2}\right)^{-2}$ (${\rm GeV^4}$)&decay channel& $Br\times\left(\frac{s_\Delta h_{ij}}{m_\Delta^2}\right)^{-2}$ (${\rm GeV^4}$)\\ \hline
{\phantom{\Large{l}}}\raisebox{+.2cm}{\phantom{\Large{j}}}
$\bar B^0\rightarrow \rho^+\rho^+e^-e^-$&$4.40\times 10^{-15}$&$\bar B^0\rightarrow \rho^+D_s^{\ast+}e^-\mu^-$&$6.67\times 10^{-16}$\\
{\phantom{\Large{l}}}\raisebox{+.2cm}{\phantom{\Large{j}}}
$\bar B^0\rightarrow \rho^+\rho^+\mu^-\mu^-$&$4.73\times 10^{-15}$&$\bar B^0\rightarrow K^{\ast+}D^{\ast+}e^-e^-$&$3.82\times 10^{-14}$\\
{\phantom{\Large{l}}}\raisebox{+.2cm}{\phantom{\Large{j}}}
$\bar B^0\rightarrow \rho^+\rho^+e^-\mu^-$&$9.09\times 10^{-15}$&$\bar B^0\rightarrow K^{\ast+}D^{\ast+}\mu^-\mu^-$&$4.11\times 10^{-14}$\\
{\phantom{\Large{l}}}\raisebox{+.2cm}{\phantom{\Large{j}}}
$\bar B^0\rightarrow \rho^+K^{\ast+}e^-e^-$&$1.46\times 10^{-16}$&$\bar B^0\rightarrow K^{\ast+}D^{\ast+}e^-\mu^-$&$7.86\times 10^{-14}$\\
{\phantom{\Large{l}}}\raisebox{+.2cm}{\phantom{\Large{j}}}
$\bar B^0\rightarrow \rho^+K^{\ast+}\mu^-\mu^-$&$1.57\times 10^{-16}$&$\bar B^0\rightarrow D^{\ast+}D^{\ast+}e^-e^-$&$6.63\times 10^{-14}$\\
{\phantom{\Large{l}}}\raisebox{+.2cm}{\phantom{\Large{j}}}
$\bar B^0\rightarrow \rho^+K^{\ast+}e^-\mu^-$&$3.01\times 10^{-16}$&$\bar B^0\rightarrow D^{\ast+}D^{\ast+}\mu^-\mu^-$&$7.24\times 10^{-14}$\\
{\phantom{\Large{l}}}\raisebox{+.2cm}{\phantom{\Large{j}}}
$\bar B^0\rightarrow \rho^+D^{\ast+}e^-e^-$&$6.61\times 10^{-13}$&$\bar B^0\rightarrow D^{\ast+}D^{\ast+}e^-\mu^-$&$1.36\times 10^{-13}$\\
{\phantom{\Large{l}}}\raisebox{+.2cm}{\phantom{\Large{j}}}
$\bar B^0\rightarrow \rho^+D^{\ast+}\mu^-\mu^-$&$7.09\times 10^{-13}$&$\bar B^0\rightarrow D^{\ast+}D_s^{\ast+}e^-e^-$&$6.69\times 10^{-13}$\\
{\phantom{\Large{l}}}\raisebox{+.2cm}{\phantom{\Large{j}}}
$\bar B^0\rightarrow \rho^+D^{\ast+}e^-\mu^-$&$1.36\times 10^{-12}$&$\bar B^0\rightarrow D^{\ast+}D_s^{\ast+}\mu^-\mu^-$&$7.29\times 10^{-13}$\\
{\phantom{\Large{l}}}\raisebox{+.2cm}{\phantom{\Large{j}}}
$\bar B^0\rightarrow \rho^+D_s^{\ast+}e^-e^-$&$3.21\times 10^{-16}$&$\bar B^0\rightarrow D^{\ast+}D_s^{\ast+}e^-\mu^-$&$1.36\times 10^{-12}$\\
{\phantom{\Large{l}}}\raisebox{+.2cm}{\phantom{\Large{j}}}
$\bar B^0\rightarrow \rho^+D_s^{\ast+}\mu^-\mu^-$&$3.51\times 10^{-16}$&&\\
\hline\hline
\end{tabular*}
\end{table}

\begin{table}
\caption{$Br\times\left(\frac{s_\Delta h_{ij}}{m_\Delta^2}\right)^{-2}$ for $0^-0^-$ decay channels of $\bar B_s^0$.}
\vspace{0.2cm}
\setlength{\tabcolsep}{0.01cm}
\centering
\begin{tabular*}{\textwidth}{@{}@{\extracolsep{\fill}}cccc}
\hline\hline
decay channel&$Br\times\left(\frac{s_\Delta h_{ij}}{m_\Delta^2}\right)^{-2}$ (${\rm GeV^4}$)&decay channel& $Br\times\left(\frac{s_\Delta h_{ij}}{m_\Delta^2}\right)^{-2}$ (${\rm GeV^4}$)\\ \hline
{\phantom{\Large{l}}}\raisebox{+.2cm}{\phantom{\Large{j}}}
$\bar B_s^0\rightarrow \pi^+K^+e^-e^-$&$6.27\times 10^{-13}$&$\bar B_s^0\rightarrow K^+D_s^+e^-\mu^-$&$1.60\times 10^{-12}$\\
{\phantom{\Large{l}}}\raisebox{+.2cm}{\phantom{\Large{j}}}
$\bar B_s^0\rightarrow \pi^+K^+\mu^-\mu^-$&$6.71\times 10^{-13}$&$\bar B_s^0\rightarrow K^+D^+e^-e^-$&$6.26\times 10^{-14}$\\
{\phantom{\Large{l}}}\raisebox{+.2cm}{\phantom{\Large{j}}}
$\bar B_s^0\rightarrow \pi^+K^+e^-\mu^-$&$1.29\times 10^{-12}$&$\bar B_s^0\rightarrow K^+D^+\mu^-\mu^-$&$6.77\times 10^{-14}$\\
{\phantom{\Large{l}}}\raisebox{+.2cm}{\phantom{\Large{j}}}
$\bar B_s^0\rightarrow K^+K^+e^-e^-$&$8.76\times 10^{-14}$&$\bar B_s^0\rightarrow K^+D^+e^-\mu^-$&$1.30\times 10^{-13}$\\
{\phantom{\Large{l}}}\raisebox{+.2cm}{\phantom{\Large{j}}}
$\bar B_s^0\rightarrow K^+K^+\mu^-\mu^-$&$9.40\times 10^{-14}$&$\bar B_s^0\rightarrow D^+D_s^+e^-e^-$&$9.22\times 10^{-13}$\\
{\phantom{\Large{l}}}\raisebox{+.2cm}{\phantom{\Large{j}}}
$\bar B_s^0\rightarrow K^+K^+e^-\mu^-$&$1.81\times 10^{-13}$&$\bar B_s^0\rightarrow D^+D_s^+\mu^-\mu^-$&$1.04\times 10^{-12}$\\
{\phantom{\Large{l}}}\raisebox{+.2cm}{\phantom{\Large{j}}}
$\bar B_s^0\rightarrow \pi^+D_s^+e^-e^-$&$1.02\times 10^{-11}$&$\bar B_s^0\rightarrow D^+D_s^+e^-\mu^-$&$1.94\times 10^{-12}$\\
{\phantom{\Large{l}}}\raisebox{+.2cm}{\phantom{\Large{j}}}
$\bar B_s^0\rightarrow \pi^+D_s^+\mu^-\mu^-$&$1.11\times 10^{-11}$&$\bar B_s^0\rightarrow D_s^+D_s^+e^-e^-$&$5.15\times 10^{-11}$\\
{\phantom{\Large{l}}}\raisebox{+.2cm}{\phantom{\Large{j}}}
$\bar B_s^0\rightarrow \pi^+D_s^+e^-\mu^-$&$2.12\times 10^{-11}$&$\bar B_s^0\rightarrow D_s^+D_s^+\mu^-\mu^-$&$5.84\times 10^{-11}$\\
{\phantom{\Large{l}}}\raisebox{+.2cm}{\phantom{\Large{j}}}
$\bar B_s^0\rightarrow K^+D_s^+e^-e^-$&$7.75\times 10^{-13}$&$\bar B_s^0\rightarrow D_s^+D_s^+e^-\mu^-$&$1.08\times 10^{-10}$\\
{\phantom{\Large{l}}}\raisebox{+.2cm}{\phantom{\Large{j}}}
$\bar B_s^0\rightarrow K^+D_s^+\mu^-\mu^-$&$8.33\times 10^{-13}$&&\\
\hline\hline
\end{tabular*}
\end{table}

\begin{table}
\caption{$Br\times\left(\frac{s_\Delta h_{ij}}{m_\Delta^2}\right)^{-2}$ for $0^-1^-$ decay channels of $\bar B_s^0$.}
\vspace{0.2cm}
\setlength{\tabcolsep}{0.01cm}
\centering
\begin{tabular*}{\textwidth}{@{}@{\extracolsep{\fill}}cccc}
\hline\hline
decay channel&$Br\times\left(\frac{s_\Delta h_{ij}}{m_\Delta^2}\right)^{-2}$ (${\rm GeV^4}$)&decay channel& $Br\times\left(\frac{s_\Delta h_{ij}}{m_\Delta^2}\right)^{-2}$ (${\rm GeV^4}$)\\ \hline
{\phantom{\Large{l}}}\raisebox{+.2cm}{\phantom{\Large{j}}}
$\bar B_s^0\rightarrow \pi^+K^{\ast+}e^-e^-$&$5.08\times 10^{-16}$&$\bar B_s^0\rightarrow K^{\ast+}D_s^{+}e^-e^-$&$1.00\times 10^{-12}$\\
{\phantom{\Large{l}}}\raisebox{+.2cm}{\phantom{\Large{j}}}
$\bar B_s^0\rightarrow \pi^+K^{\ast+}\mu^-\mu^-$&$5.44\times 10^{-16}$&$\bar B_s^0\rightarrow K^{\ast+}D_s^{+}\mu^-\mu^-$&$1.09\times 10^{-12}$\\
{\phantom{\Large{l}}}\raisebox{+.2cm}{\phantom{\Large{j}}}
$\bar B_s^0\rightarrow \pi^+K^{\ast+}e^-\mu^-$&$1.05\times 10^{-15}$&$\bar B_s^0\rightarrow K^{\ast+}D_s^{+}e^-\mu^-$&$2.09\times 10^{-12}$\\
{\phantom{\Large{l}}}\raisebox{+.2cm}{\phantom{\Large{j}}}
$\bar B_s^0\rightarrow K^+K^{\ast+}e^-e^-$&$6.94\times 10^{-14}$&$\bar B_s^0\rightarrow K^+D^{\ast+}e^-e^-$&$6.19\times 10^{-14}$\\
{\phantom{\Large{l}}}\raisebox{+.2cm}{\phantom{\Large{j}}}
$\bar B_s^0\rightarrow K^+K^{\ast+}\mu^-\mu^-$&$7.44\times 10^{-14}$&$\bar B_s^0\rightarrow K^+D^{\ast+}\mu^-\mu^-$&$6.66\times 10^{-14}$\\
{\phantom{\Large{l}}}\raisebox{+.2cm}{\phantom{\Large{j}}}
$\bar B_s^0\rightarrow K^+K^{\ast+}e^-\mu^-$&$1.43\times 10^{-13}$&$\bar B_s^0\rightarrow K^+D^{\ast+}e^-\mu^-$&$1.28\times 10^{-13}$\\
{\phantom{\Large{l}}}\raisebox{+.2cm}{\phantom{\Large{j}}}
$\bar B_s^0\rightarrow \rho^+K^{+}e^-e^-$&$1.28\times 10^{-12}$&$\bar B_s^0\rightarrow K^{\ast+}D^{+}e^-e^-$&$2.79\times 10^{-17}$\\
{\phantom{\Large{l}}}\raisebox{+.2cm}{\phantom{\Large{j}}}
$\bar B_s^0\rightarrow \rho^+K^{+}\mu^-\mu^-$&$1.37\times 10^{-12}$&$\bar B_s^0\rightarrow K^{\ast+}D^{+}\mu^-\mu^-$&$3.02\times 10^{-17}$\\
{\phantom{\Large{l}}}\raisebox{+.2cm}{\phantom{\Large{j}}}
$\bar B_s^0\rightarrow \rho^+K^{+}e^-\mu^-$&$2.64\times 10^{-12}$&$\bar B_s^0\rightarrow K^{\ast+}D^{+}e^-\mu^-$&$5.78\times 10^{-17}$\\
{\phantom{\Large{l}}}\raisebox{+.2cm}{\phantom{\Large{j}}}
$\bar B_s^0\rightarrow \pi^+D_s^{\ast+}e^-e^-$&$1.72\times 10^{-13}$&$\bar B_s^0\rightarrow D_s^+D_s^{\ast+}e^-e^-$&$8.29\times 10^{-12}$\\
{\phantom{\Large{l}}}\raisebox{+.2cm}{\phantom{\Large{j}}}
$\bar B_s^0\rightarrow \pi^+D_s^{\ast+}\mu^-\mu^-$&$1.81\times 10^{-13}$&$\bar B_s^0\rightarrow D_s^+D_s^{\ast+}\mu^-\mu^-$&$9.22\times 10^{-12}$\\
{\phantom{\Large{l}}}\raisebox{+.2cm}{\phantom{\Large{j}}}
$\bar B_s^0\rightarrow \pi^+D_s^{\ast+}e^-\mu^-$&$3.50\times 10^{-13}$&$\bar B_s^0\rightarrow D_s^+D_s^{\ast+}e^-\mu^-$&$1.72\times 10^{-11}$\\
{\phantom{\Large{l}}}\raisebox{+.2cm}{\phantom{\Large{j}}}
$\bar B_s^0\rightarrow \rho^+D_s^{+}e^-e^-$&$1.92\times 10^{-11}$&$\bar B_s^0\rightarrow D^+D_s^{\ast+}e^-e^-$&$4.78\times 10^{-14}$\\
{\phantom{\Large{l}}}\raisebox{+.2cm}{\phantom{\Large{j}}}
$\bar B_s^0\rightarrow \rho^+D_s^{+}\mu^-\mu^-$&$2.09\times 10^{-11}$&$\bar B_s^0\rightarrow D^+D_s^{\ast+}\mu^-\mu^-$&$5.01\times 10^{-14}$\\
{\phantom{\Large{l}}}\raisebox{+.2cm}{\phantom{\Large{j}}}
$\bar B_s^0\rightarrow \rho^+D_s^{+}e^-\mu^-$&$3.99\times 10^{-11}$&$\bar B_s^0\rightarrow D^+D_s^{\ast+}e^-\mu^-$&$9.56\times 10^{-14}$\\
{\phantom{\Large{l}}}\raisebox{+.2cm}{\phantom{\Large{j}}}
$\bar B_s^0\rightarrow K^+D_s^{\ast+}e^-e^-$&$1.15\times 10^{-12}$&$\bar B_s^0\rightarrow D^{\ast+}D_s^{+}e^-e^-$&$4.86\times 10^{-13}$\\
{\phantom{\Large{l}}}\raisebox{+.2cm}{\phantom{\Large{j}}}
$\bar B_s^0\rightarrow K^+D_s^{\ast+}\mu^-\mu^-$&$1.24\times 10^{-12}$&$\bar B_s^0\rightarrow D^{\ast+}D_s^{+}\mu^-\mu^-$&$5.39\times 10^{-13}$\\
{\phantom{\Large{l}}}\raisebox{+.2cm}{\phantom{\Large{j}}}
$\bar B_s^0\rightarrow K^+D_s^{\ast+}e^-\mu^-$&$2.38\times 10^{-12}$&$\bar B_s^0\rightarrow D^{\ast+}D_s^{+}e^-\mu^-$&$1.01\times 10^{-11}$\\
\hline\hline
\end{tabular*}
\end{table}

\begin{table}
\caption{$Br\times\left(\frac{s_\Delta h_{ij}}{m_\Delta^2}\right)^{-2}$ for $1^-1^-$ decay channels of $\bar B_s^0$.}
\vspace{0.2cm}
\setlength{\tabcolsep}{0.01cm}
\centering
\begin{tabular*}{\textwidth}{@{}@{\extracolsep{\fill}}cccc}
\hline\hline
decay channel&$Br\times\left(\frac{s_\Delta h_{ij}}{m_\Delta^2}\right)^{-2}$ (${\rm GeV^4}$)&decay channel& $Br\times\left(\frac{s_\Delta h_{ij}}{m_\Delta^2}\right)^{-2}$ (${\rm GeV^4}$)\\ \hline
{\phantom{\Large{l}}}\raisebox{+.2cm}{\phantom{\Large{j}}}
$\bar B_s^0\rightarrow \rho^+K^{\ast+}e^-e^-$&$1.54\times 10^{-15}$&$\bar B_s^0\rightarrow K^{\ast+}D_s^{\ast+}e^-\mu^-$&$1.30\times 10^{-13}$\\
{\phantom{\Large{l}}}\raisebox{+.2cm}{\phantom{\Large{j}}}
$\bar B_s^0\rightarrow \rho^+K^{\ast+}\mu^-\mu^-$&$1.66\times 10^{-15}$&$\bar B_s^0\rightarrow K^{\ast+}D^{\ast+}e^-e^-$&$2.07\times 10^{-16}$\\
{\phantom{\Large{l}}}\raisebox{+.2cm}{\phantom{\Large{j}}}
$\bar B_s^0\rightarrow \rho^+K^{\ast+}e^-\mu^-$&$3.19\times 10^{-15}$&$\bar B_s^0\rightarrow K^{\ast+}D^{\ast+}\mu^-\mu^-$&$2.26\times 10^{-16}$\\
{\phantom{\Large{l}}}\raisebox{+.2cm}{\phantom{\Large{j}}}
$\bar B_s^0\rightarrow K^{\ast+}K^{\ast+}e^-e^-$&$1.70\times 10^{-16}$&$\bar B_s^0\rightarrow K^{\ast+}D^{\ast+}e^-\mu^-$&$4.30\times 10^{-16}$\\
{\phantom{\Large{l}}}\raisebox{+.2cm}{\phantom{\Large{j}}}
$\bar B_s^0\rightarrow K^{\ast+}K^{\ast+}\mu^-\mu^-$&$1.84\times 10^{-16}$&$\bar B_s^0\rightarrow D^{\ast+}D_s^{\ast+}e^-e^-$&$3.07\times 10^{-14}$\\
{\phantom{\Large{l}}}\raisebox{+.2cm}{\phantom{\Large{j}}}
$\bar B_s^0\rightarrow K^{\ast+}K^{\ast+}e^-\mu^-$&$3.52\times 10^{-16}$&$\bar B_s^0\rightarrow D^{\ast+}D_s^{\ast+}\mu^-\mu^-$&$3.36\times 10^{-14}$\\
{\phantom{\Large{l}}}\raisebox{+.2cm}{\phantom{\Large{j}}}
$\bar B_s^0\rightarrow \rho^+D_s^{\ast+}e^-e^-$&$5.69\times 10^{-13}$&$\bar B_s^0\rightarrow D^{\ast+}D_s^{\ast+}e^-\mu^-$&$6.29\times 10^{-14}$\\
{\phantom{\Large{l}}}\raisebox{+.2cm}{\phantom{\Large{j}}}
$\bar B_s^0\rightarrow \rho^+D_s^{\ast+}\mu^-\mu^-$&$6.12\times 10^{-13}$&$\bar B_s^0\rightarrow D_s^{\ast+}D_s^{\ast+}e^-e^-$&$1.23\times 10^{-12}$\\
{\phantom{\Large{l}}}\raisebox{+.2cm}{\phantom{\Large{j}}}
$\bar B_s^0\rightarrow \rho^+D_s^{\ast+}e^-\mu^-$&$1.17\times 10^{-12}$&$\bar B_s^0\rightarrow D_s^{\ast+}D_s^{\ast+}\mu^-\mu^-$&$1.35\times 10^{-12}$\\
{\phantom{\Large{l}}}\raisebox{+.2cm}{\phantom{\Large{j}}}
$\bar B_s^0\rightarrow K^{\ast+}D_s^{\ast+}e^-e^-$&$6.29\times 10^{-14}$&$\bar B_s^0\rightarrow D_s^{\ast+}D_s^{\ast+}e^-\mu^-$&$2.51\times 10^{-12}$\\
{\phantom{\Large{l}}}\raisebox{+.2cm}{\phantom{\Large{j}}}
$\bar B_s^0\rightarrow K^{\ast+}D_s^{\ast+}\mu^-\mu^-$&$6.81\times 10^{-14}$&&\\
\hline\hline
\end{tabular*}
\end{table}

\section{Conclusions}

As a conclusion, we have studied the lepton number violation four-body decay processes of neutral flavored mesons, including $\bar K^0$, $D^0$, $\bar B^0$, and $\bar B_s^0$. They are assumed to be induced by a doubly-charged scalar. For $\bar K^0$, the channel $\bar K^0\rightarrow \pi^+\pi^+e^-e^-$ has the largest value of $Br\times\left(\frac{s_\Delta h_{ij}}{m_\Delta^2}\right)^{-2}$, which is of the order of $10^{-11}$ ${\rm GeV^4}$. For $D^0$, the channel $D^0\rightarrow \pi^-K^-l_1^+l_2^+$ has the largest order of magnitude $10^{-10}$ ${\rm GeV^4}$. By comparing with the E791 experimental data, we set the upper limit for $\frac{s_\Delta h_{ij}}{m_\Delta^2}$ being of the order of $10^3$ ${\rm GeV^{-2}}$. For $\bar B^0$ and $\bar B_s^0$, the largest value of $Br\times\left(\frac{s_\Delta h_{ij}}{m_\Delta^2}\right)^{-2}$ is also about $10^{-10}$ ${\rm GeV^4}$. These results may provide some help for the studies of neutrinoless double beta decays of mesons. We expect more experimental detections of such processes by the LHCb and Belle-II Collaborations. 

\section*{ACKNOWLEDGEMENTS}
This work was supported in part by the National Natural Science
Foundation of China (NSFC) under Grants No.~11405037, No.~11575048, and No.~11505039.

\appendix
\section{Wave functions of mesons}

With the instantaneous approximation, the Bethe-Salpeter wave function of the meson fulfills the full Salpeter equations~\cite{Kim}
\begin{equation}
\label{salpeter}
\begin{aligned}
&(M-\omega_1-\omega_2)\varphi_{_P}^{++}(q_\perp) = \Lambda_1^+\eta_{_P}(q_\perp)\Lambda_2^+,\\
&(M+\omega_1+\omega_2)\varphi_{_P}^{--}(q_\perp) = - \Lambda_1^-\eta_{_P}(q_\perp)\Lambda_2^-,\\ &\varphi_{_P}^{+-}(q_\perp) = \varphi_{_P}^{-+}(q_\perp) = 0,
\end{aligned}
\end{equation}
where $q^\mu_\perp=q^\mu-\frac{P\cdot q}{M^2}P^\mu$, $\omega_1=\sqrt{m_1^2-q_\perp^2}$, and $\omega_2=\sqrt{m_2^2-q_\perp^2}$; $m_1$ and $m_2$ are the masses of quark and antiquark, respectively; $\Lambda_i^{\pm}=\frac{1}{2\omega_i}\left\{\frac{\slashed P}{M}\omega_i\pm\left[\slashed q_\perp + (-1)^{i+1} m_i\right]\right\}$ is the projection operator. In the above equation, we have defined 
\begin{equation}
\label{eta}
\eta_{_P}(q_\perp)=\int\frac{d^3k_\perp}{(2\pi)^3} V(P; q_\perp,k_\perp)\varphi_{_P}(k_\perp),
\end{equation}
and 
\begin{equation}
\varphi_{_P}^{\pm\pm}(q_\perp)=\Lambda^{\pm}_1\frac{\slashed P}{M}\varphi_{_P}(q_\perp)\frac{\slashed P}{M}\Lambda^{\pm}_2,
\end{equation} 
with $\varphi_{_P}(q_\perp)$ being the wave function, which is constructed by $\slashed q_\perp$, $\slashed P$, and polarization vector.  Here we only present the expression for the positive energy part of the wave function. For the $1^-$ state, it has the form
\begin{equation}
\begin{aligned}
\varphi^{++}_{1^{-}}(q_\perp)&=(q_\perp\cdot\epsilon)\left[A_1(q_\perp)+\frac{\slashed{P}}{M}A_2(q_\perp)
+\frac{\slashed{q}_\perp}{M}A_3(q_\perp)+\frac{\slashed{P}\slashed{q}_\perp}{M^2}A_4(q_\perp)\right]\\
&~~~~+ M\slashed\epsilon\left[A_5(q_\perp)+\frac{\slashed{P}}{M}A_6(q_\perp)
+\frac{\slashed{q}_\perp}{M}A_7(q_\perp)+\frac{\slashed{P}\slashed{q}_\perp}{M^2}A_8(q_\perp)\right].
\end{aligned}
\end{equation}
For the $0^-$ state, it has the form
\begin{equation}
\begin{aligned}
\varphi^{++}_{0^{-}}(q_\perp)&=\left[B_1(q_{\perp})+\frac{\slashed{P}}{M}B_2(q_{\perp})
+\frac{\slashed{q}_{\perp}}{M}B_3(q_{\perp})+\frac{\slashed{P}\slashed{q}_{\perp}}{M^2}B_4(q_{\perp})\right]\gamma_5.
\end{aligned}
\end{equation}
$A_i$ and $B_i$ are functions of $q_\perp^2$, whose numerical results are achieved by solving Eq.~(A1).

The interaction potential used in this work has the form~\cite{Kim}
\begin{equation}
\begin{aligned}
\label{Cornell}
V({\vec q})=V_s(\vec{q})
+\gamma_0\otimes\gamma^0V_v(\vec{q}),
\end{aligned}
\end{equation}
where
\begin{equation}
\begin{aligned}
&V_{s}(\vec{q})
=-\left(\frac{\lambda}{\alpha}+V_0\right)\delta^{3}(\vec{q})
+\frac{\lambda}{\pi^{2}}\frac{1}{(\vec{q}^{2}+\alpha^{2})^{2}},\\
&V_v(\vec{q})=-\frac{2}{3\pi^{2}}
\frac{\alpha_{s}(\vec{q})}{\vec{q}^{2}+\alpha^{2}},\\
&\alpha_s(\vec{q})=\frac{12\pi}{27}
\frac{1}{{\rm{ln}}\left(a+\frac{\vec{q}^2}{\Lambda^2_{QCD}}\right)}.
\end{aligned}
\end{equation}
The parameters involved are $a=e=2.71828$, $\alpha=0.06$ GeV, $\lambda=0.21$ ${\rm GeV}^2$, $\Lambda_{QCD}=0.27$ GeV; $V_0$ is decided by fitting the mass of the ground state. The constituent quark masses used here are $m_b=4.96$ GeV, $m_c=1.62$ GeV, $m_s=0.5$ GeV, $m_u=0.305$ GeV, and $m_d=0.311$ GeV.

%%%%%%%%%%%%%%%%%%%%%%%%%%%%%%%%%%

\end{document}